# Optimal chemotactic navigation in disordered landscapes

Yang BAI[1,2*+], Caiyun HE[1,2*], Weirong LIU[1,2], Songtao Cheng[1,2], Pan Chu[1,2], Liang Luo[3], Chenli Liu[1,2], Xiongfei FU[1,2+]

[1]State Key Laboratory for Quantitative Synthetic Biology, Shenzhen Institute of Synthetic Biology, Shenzhen Institutes of Advanced Technology, Chinese Academy of Sciences, Shenzhen, 518055, China

[2]University of Chinese Academy of Sciences, Beijing, 101408, China

[3]Huazhong Agricultural University, Wuhan, 430070, China

*Yang BAI & Caiyun HE contributed equally to this work.

+To whom correspondence should be addressed

**Email:** xiongfei.fu@siat.ac.cn and yang.bai@siat.ac.cn

## Abstract

Active navigation in disordered landscape with random traps arises from the interplay between directed motion and environmental constraints. Using *E. coli* chemotactic navigation in agar gels as a model system, we reveal a fundamental trade-off between diffusive exploration and chemotactic directional bias that dictates the optimal strategy for population range expansion. Counter-intuitively, evolution selects for shorter mean run times ($\tau_f$) to achieve faster chemotactic migration in denser environments. Controlled experiments reveal a non-monotonic relationship between chemotactic navigation speed and $\tau_f$, with the optimum shifting according to the density of physical traps in the gel. Single-cell analysis indicates that cells primarily escape a trap without an intervening tumble and without a large reorientation. Based on these observations, we develop a minimal theoretical model showing that the optimal $\tau_f$ emerges from an antagonistic scaling: while the diffusion coefficient increases with $\tau_f$, the chemotactic bias coefficient decreases with it. This work establishes an optimality condition for biased run-tumble motion in disordered landscapes with random traps.

## Introduction

Bacteria navigate dynamic environments by actively modulating their run-and-tumble motility, a fundamental mechanism for sensing and responding to chemical gradients, known as chemotaxis[1-8]. This locomotion alternates between directed movement ("runs"), propelled by rotating flagellar bundles, and stochastic reorientation ("tumbles"), which randomize the cell's direction[9-15]. By adjusting run duration in response to temporal changes in attractant concentration, bacterial cells bias their motion to migrate directionally along chemoattractant gradients[9,16-19].

In homogeneous liquid environments, the chemotactic ability ($\chi$) of a bacterium is often proportional to its translational diffusion coefficient ($D$) [1,20-23], which is intrinsically set by the mean free runtime, $\tau_f$. However, in natural habitats such as soil, porous gels, or mucous layers, bacteria encounter disordered landscapes where physical obstacles act as transient traps[24-27]. These obstacles confine motility, introducing an external trapping timescale, $\tau_t$, that operates alongside the intrinsic tumbling timescale. While theoretical and experimental studies have examined how this interplay affects diffusive transport[24,26,28-30], a critical question remains: how does it fundamentally shape chemotactic searching and navigation strategies?

Understanding this interplay is critical, both for single-cell dynamics and for population-level adaptation. Bacterial groups can dynamically generate their own chemoattractant gradients by metabolizing local resources, facilitating coordinated migration and rapid colonization that outstrips classical growth-diffusion dynamics[1,7]. This capability confers a significant evolutionary advantage. Consequently, a central question arises: how do bacteria optimize their intrinsic motility parameters, such as $\tau_f$, to maximize navigation efficiency under external constraints of a disordered landscape (Fig. S1)? Understanding this adaptive optimization is key to deciphering microbial ecology and holds promise for applications in bioremediation and synthetic biology.

In this work, we investigate how bacteria optimize chemotactic navigation in disordered landscapes. Through experimental evolution, we identified a density-dependent optimal mean free run time ($\tau_f^{opt}$) that maximizes the population chemotactic navigation speed. Using a strain with titratable $\tau_f$, we demonstrate a non-monotonic relationship between the chemotactic navigation speed and $\tau_f$, where $\tau_f^{opt}$ shifts according to environmental trap density. Single-cell tracking reveals that cells primarily escape a trap without an intervening tumble and without a large reorientation. Motivated by this observation, we developed a theoretical model showing that while prolonged runs enhance diffusion, they paradoxically suppress chemotactic bias in disordered landscape. This trade-off between directional bias and diffusive exploration explains the observed non-monotonic dependence of chemotactic performance on $\tau_f$. Our findings elucidate a key adaptive strategy by which bacteria tune their motility, balancing exploration and trapping to navigate disordered landscape.

## Results

### Phenotypic evolution of chemotactic bacteria in agar gels

To understand how chemotactic bacteria adapt to disordered landscapes, we performed a spatial evolution experiment selecting for rapid range expansion in *E. coli* MG1655 populations[31-33]. We propagated populations on semi-solid agar plates at three different concentrations (0.2% 0.25% 0.3%), which modulate the gel's pore size and density, thereby creating random traps that impede bacterial motility[34]. Bacteria were inoculated at the center of each plate; nutrient consumption generated self-generated chemoattractant gradients that drove outward migration. After 24 hours—sufficient for full plate colonization—we transferred cells from the migration front to a fresh plate, repeating this selection process over 40 cycles (~500 generations) (Fig. 1a).

We quantified the chemotactic navigation by imaging the expansion front, which advanced linearly with time. Over successive cycles, phenotypically evolved populations selected in the expansion front at all three agar concentrations exhibited a progressive increase in chemotactic navigation (Fig. 1b). Crucially, the growth rate of the selected populations, measured every 5 cycles, remained constant throughout the experiment (see Methods and Fig. S2a). Given that chemotactic navigation is governed by the chemotaxis coefficient ($\chi$) when growth rate is unchanged[1], these results indicate that selection specifically enhanced chemotactic performance in these obstructed environments.

To identify the mechanistic basis of this adaptation, we analyzed single-cell trajectories of evolved populations using a customized tracking platform[8,18,19,35]. This revealed agar concentration-dependent shifts in motility parameters. The mean free run time ($\tau_f$) and mean free run length of evolved cells decreases as the agar concentration increases from 0.2% to 0.3% (Fig. 1c, Fig. S2b). Notably, the tumble duration ($\tau_{tmb}$) remained constant while the mean run speed ($v_0$) increases slightly during evolution (~15%, Fig. S2c, d), and tumble bias (defined by fraction of time spent tumbling, independent of the turning angle) increased accordingly (Fig. S2e). Key motility parameters of the evolved strains including run times, run lengths, and tumble durations, retained Poisson-distributed dynamics after 40 evolutionary cycles (Fig. S3a-c). Meanwhile, tumble bias and run speed exhibited unimodal distributions after evolution (Fig. S3 d,e), indicating that evolutionary tuning of $\tau_f$ occurred without destabilizing the core chemotaxis regulatory network[22,23].

Together, these findings demonstrate that adaptation to disordered landscapes involves the precise modulation of the intrinsic run time, $\tau_f$. This evolutionary tuning optimizes motility by balancing the exploratory benefit of long runs against the increased risk of trapping. This raises a central question: how does environmental trap density mechanistically define the optimal mean free run time for efficient chemotactic navigation?

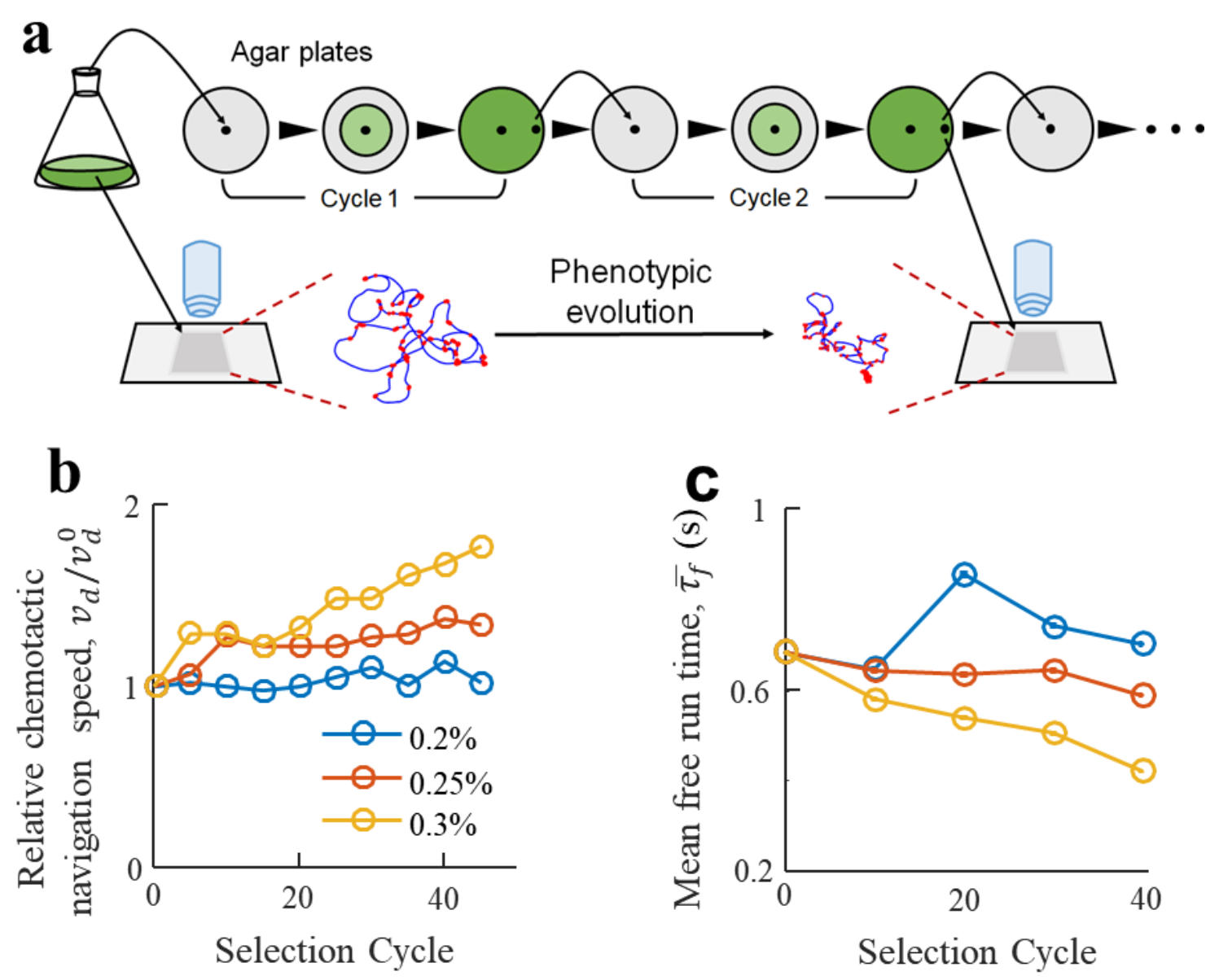

Figure 1. **Phenotypic evolution of bacterial chemotaxis in disordered landscapes**. (a) Schematic of the experimental evolution protocol for selecting rapid range expansion in semi-solid agar gels. Bacterial populations are serially passage through agar plates in alternating selection cycles with periodic sampling and phenotypic characterization. (b) The relative chemotactic navigation speed ($v_d/v_d^0$), normalized to that of the ancestor cells, increased over time across all three selection intensities (0.2%, 0.25%, and 0.3% agar), indicating enhanced motility under stronger selection. The chemotactic navigation speed ($v_d$ ) was determined by fitting colony radii from time-lapse backlit agar plate images (see Methods). Error bars (n = 3) are smaller than the markers. (c) The mean free run time $\tau_f$ of selected cells measured in liquid medium changed during selection cycles and converged to distinct, agar concentration-dependent values. Cells were imaged and tracked using U-Track, with motion classified into run and tumble states via clustering based on speed, acceleration, and angular velocity; $\tau_f$ was then calculated as the average duration of run events (see Methods). Each measurement included over 100,000 run events. Error bars (SEM) are smaller than the symbol size.

## Non-monotonic dependence of chemotactic navigation on run time

The evolutionary tuning of $\tau_f$ suggests the existence of an optimal free runtime $\tau_f^{opt}$ for bacterial chemotactic navigation in disordered landscapes. To test this hypothesis, we used a $\tau_f$ titratable strain in which $\tau_f$ increases linearly with the logarithm of the external inducer anhydrotetracycline (aTc) concentration (Fig. 2a, Fig. S4 & Methods). Using this strain, we measured the speed of population chemotactic navigation ($V_d$) in agar gels of 0.2% and 0.3%. We observed a pronounced non-monotonic dependence of $V_d$ on $\tau_f$: the speed initially increased with longer run times but declined after surpassing a distinct optimum $\tau_f^{opt}$(Fig.2b). This non-monotonic dependence implies a fundamental trade-off: while prolonged runs enhance diffusion exploration and gradient sensing in open environments, they concurrently increase the probability of becoming trapped in a disordered landscape that penetrates chemotactic navigation.

Critically, the value of $\tau_f^{opt}$ is itself dependent on the properties of the environment. A key implication of this dependency is that a given pair of strains will display divergent spatial patterns when cultured on gels of varying densities. Competition assays confirmed this environmental sensitivity, revealing that $\tau_f^{opt}$ is significantly smaller in denser agar (0.3%) than in softer (0.2%) agar (Fig. S5). This inverse relationship between $\tau_f^{opt}$ and gel density allowed strains with $\tau_f^{opt}$ values closer to the environment-specific $\tau_f^{opt}$ to gain a selective advantage during range expansion. These results are in qualitative agreement with our evolutionary trajectories (Fig. 1c), confirming that selection drives populations toward the $\tau_f$ that maximizes the chemotactic navigation in a given landscape [7,33].

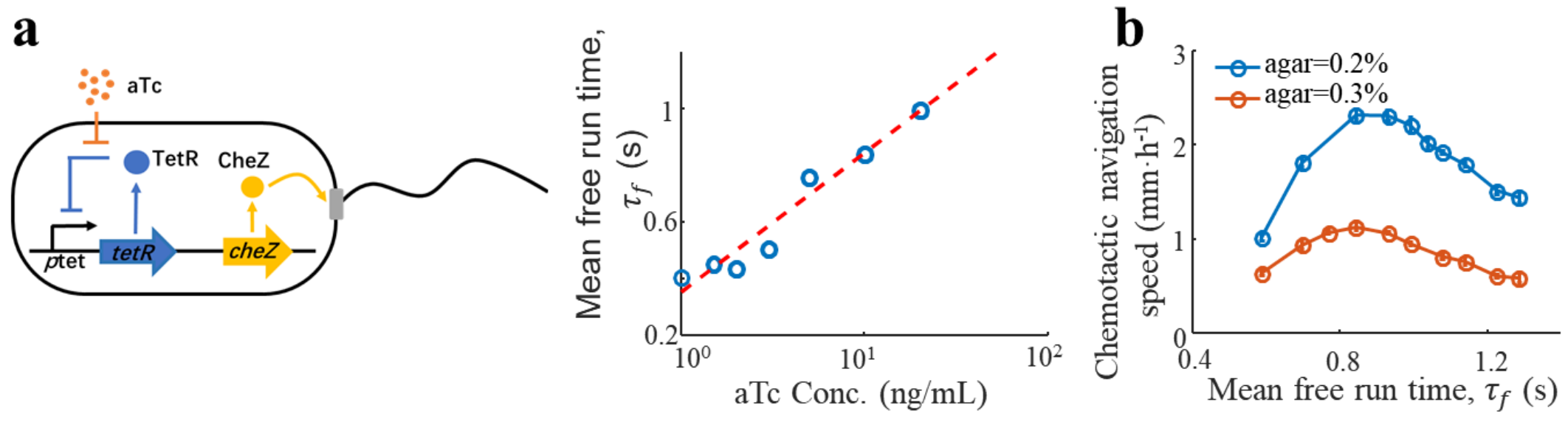

**Figure 2. Non-monotonic navigation speed reveals an optimal run time.**

(a) Design and validation of the $\tau_f$-titratable strain. Left: Genetic circuit for anhydrotetracycline (aTc)-inducible control of *cheZ* expression. Right*:* The mean free run time ($\tau_f$) of the engineered strain increases linearly with the logarithm of the aTc concentration. (dashed red line is the linear fit). More than 8000 cells were tracked at each aTc concentration, the standard error of the mean (SEM) of $\tau_f$ is smaller than marker size in all conditions. (b) Chemotactic navigation speed exhibits a non-monotonic dependence on the intrinsic free run time. Three replicates were performed for each condition with error smaller than marker size. The optimal free run time that maximizes chemotactic navigation is larger in 0.3% agar than in 0.2% agar.

**Bacterial motion with random traps**

To elucidate bacterial navigation in porous hydrogels, we sought to distinguish intrinsic behavioral states (running and tumbling) from extrinsic immobilization caused by physical confinement (trapping). In *E. coli*, running is driven by the rotation of a bundled flagellar motor, while tumbling is triggered by flagellar unbundling. We hypothesized that a third state exists: cells with bundled flagella that are physically arrested by the gel matrix. To differentiate these states, we employed an *E. coli* strain with modified FliC proteins, enabling specific fluorescent labeling of flagella[15] (Fig. 3a & Methods).

Using high-resolution time-lapse fluorescence microscopy, we simultaneously tracked flagellar dynamics and cellular trajectories (Fig. 3b, Movies S1 and S2). A custom machine-learning pipeline, based on the YOLOv5 architecture, was trained to automatically classify flagellar configurations as "bundled" or "split" states (Fig. 3c, left). Instantaneous swimming velocities were computed and normalized to the 95th percentile speed of each trajectory to account for cell-to-cell variability[8] (see Methods). This revealed a distinct bimodal velocity distribution for the bundled state in agar, contrasting with the unimodal distribution in liquid (Fig. S6a). A new peak emerged near zero velocity, corresponding to cells with active, bundled flagella whose motion was physically restrained by the gel—defining the "trapped" state (Fig. 3c, right). The speed distribution of bacteria with split flagella in liquid shows a distinct rightward shoulder, likely reflecting an intermediate semi-coiled state in which one or two flagella detach from the bundle while sustained motility is maintained. By integrating flagellar morphology classification with a dual Gaussian distribution model to differentiate between states, we resolved three distinct motility states within individual trajectories: running, tumbling, and trapping (Fig. S6b).

We next investigated the kinetics of these states. Run and tumble durations followed nearly exponential distributions, as described before[36], while trapping duration exhibited a stretched exponential distribution with a heavy tail (Fig. S6c), indicating heterogeneous escape kinetics governed by variations in local pore geometry. We quantified directional persistence by measuring the angular change between the incoming and outgoing run directions separated by a trapping or tumbling event (Fig S6d). While reorientation angles between successive runs separated by a tumble (run–tumble–run events, Fig. 3d red and Fig. S6d) were nearly isotropic, those following trap release showed a strong bias toward the original direction of motion (Fig. 3d blue).

This pronounced asymmetry reveals that the agar gel does not function as a rigid barrier. Instead, bacteria appear to navigate around or through transient, deformable traps without significantly altering their heading or relying on tumbling to execute detours. Consequently, our findings suggest a new role of tumbling in disordered landscapes: rather than serving as an essential escape mechanism, tumbling appears to play a more nuanced role in the broader chemotactic strategy within disordered landscapes.

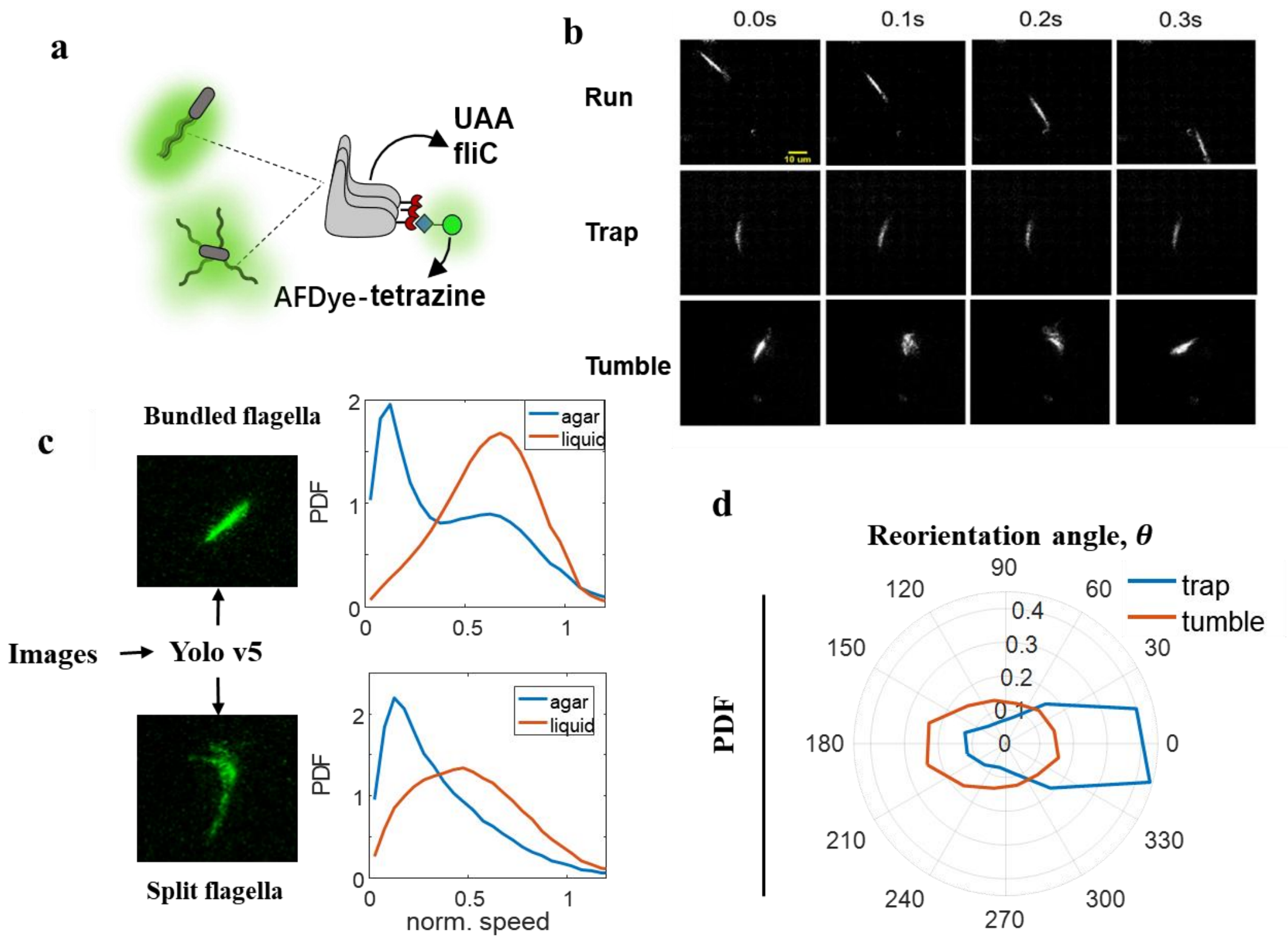

**Figure 3. Single-cell analysis of bacterial motility.**
(a) Fluorescent labeling of bacterial flagella is achieved by incorporating an unnatural amino acid into the FliC protein, followed by conjugation with AFDye-tetrazine, enables visualization of flagellar structure and dynamics. (b) Representative fluorescence time-laps micrograph of an *E. coli* cell embedded in a 0.2% agar gel, showing three distinct behavioral states: Run (cell in motion with bundled flagella); Trap (cell stopped with bundled flagella); Tumble (cell stopped with split flagella). (c) Left: Representative image frames with cells automatically identified and classified as having bundled or split flagella using an in-house trained YOLOv5 detection model. Right: Cells were tracked across frames to reconstruct trajectories, and velocities were calculated and annotated with corresponding flagellar states. The normalized run speed (normalized to the 95th percentile speed of each individual trajectory) is displayed as probability density distributions for each flagellar configuration: bundled (running or trapped) and split (tumbling), in both liquid medium (red curves) and in 0.2% agar gel (blue curves), revealing distinct motility dynamics across environments. The liquid “split” distribution may thus comprise both complete tumbles and high-speed partial-split events, biasing it toward higher speeds and plausibly explaining its overlap with the bundled distribution. (d) Probability density functions (PDF) of reorientation angles defined by changing angles between runs of successive run- tumbling-run events (red) and run-trap-run events (blue) in agar gel (Fig S6d).

**Modeling bacterial chemotaxis in disordered landscapes**

Informed by our single-cell characterization of trapping and tumbling events (Fig. 3 & Fig. S6), we model bacterial motility in disordered landscapes as a combination of two independent stochastic processes: run-tumble and run-trap dynamics (Fig. 4a). The intrinsic run-tumble process is governed by switching rates $\lambda_{rt}$ (run to tumble) and $\lambda_{tr}$ (tumble to run), which define the mean free run time $\tau_f = 1/\lambda_{rt}$ and the mean tumble duration $\tau_{tmb} = 1/\lambda_{tr}$ . Simultaneously, the extrinsic run-trap process interrupts runs at a rate $r_{rt}$ ($\tau_t = 1/r_{rt}$, mean runtime between traps) and detain cells for a mean trapping time $\tau_{trp} = 1/r_{tr}$. The effective duration of an uninterrupted run in the gel, $\tau_R$, is thus determined by the combined probability of intrinsic tumbling and extrinsic trapping: $\tau_R = \left(\tau_f^{-1} + \tau_t^{-1}\right)^{-1}$.

Assuming, for simplicity, uniform reorientation angles for both tumbling and trapping events, the diffusion coefficient *D* in a 1D system is given by: $D = \frac{v_0^2}{2}\frac{{\tau_R}^2}{\tau_R+\tau_S}$, where $v_0$ is the run speed and $\tau_S$ is the mean stop time, $\tau_S = P_{tmb}\tau_{tmb} + P_{trp}\tau_{trp}$, where $P_{tmb}$ and $P_{trp}$ defined the relative probabilities of a run terminating by tumble or by trap, with $P_{tmb} = \tau_f^{-1}\left(\tau_f^{-1} + \tau_t^{-1}\right)^{-1}$and $P_{trp} = \tau_t^{-1}\left(\tau_f^{-1} + \tau_t^{-1}\right)^{-1}$. For a fixed trapping time $\tau_{trp}$ at a given agar concentration, this model predicts the diffusion coefficient $D$ increases monotonically with $\tau_f$ (Fig. 4b). Crucially, because trapping events do not induce negative directional bias (Fig. 4d), this model alone cannot explain the non-monotonic behavior observed in our experiments, in contrast to previous models that proposed tumbling as a mechanism for bacteria to escape traps [29,37,38].

To elucidate the non-monotonic relationship between chemotactic navigation speed ($\chi$) and mean free runtime ($\tau_f$), we introduced biased runs to model chemotaxis in agar gels. Cells extend their runs when moving up chemoattractant gradients ($\tau_f^+$) and shorten them when moving down ($\tau_f^-$) (Fig. 4c). Reflecting the fundamental principle of bacterial chemotaxis with sensory adaptation[39-41], we assume run-time deviation ($\delta\tau_f$) is proportional to mean free runtime: $\delta\tau_f \equiv \tau_f^+ - \tau_f \equiv \tau_f - \tau_f^- = \alpha G \tau_f$. The parameter $\alpha$ (in unit of length) represents the biased run length of bacteria along an external chemoattractant gradient $G$ (in unit of 1/length), arising from the internal chemotactic response. Consequently, the product $\alpha G$ constitutes a dimensionless parameter. The effective free runtimes in gradient-aligned ($\pm$) directions becomes $\tau_R^{\pm} = \frac{\tau_f^{\pm}\tau_t}{\tau_f^{\pm}+\tau_t}$.

This asymmetry generates a drift velocity: $V_d = \frac{v_0 \delta\tau_R}{2(\tau_R+\tau_S)}$, where $\delta\tau_R \equiv (\tau_R^+ - \tau_R^-) \approx \frac{\alpha G \tau_f \tau_t^2}{(\tau_f+\tau_t)^2} = \frac{\alpha G}{\tau_f}\tau_R^2$, which further defines the chemotactic ability $\chi = V_d/G \approx \frac{v_0}{d}\frac{\alpha}{\tau_f}\frac{\tau_R^2}{\tau_R+\tau_S}$. This chemotactic ability $\chi$ can be decomposed into the product of a diffusion coefficient $D$ and a bias coefficient *B*: $\chi = D \cdot B$, where $B \equiv \frac{\delta\tau_R}{v_0\tau_R^2} \approx \frac{\alpha}{v_0\tau_f} = \frac{\alpha}{l_f}$, representing the relative biased run length over mean run length ($l_f$).

This formulation reveals the core trade-off: while the diffusion coefficient *D* increases with $\tau_f$ (Fig. 4b), the bias coefficient *B* decreases inversely with $\tau_f$ (Fig. 4d). Their product, chemotactic ability $\chi$, therefore exhibits a maximum at an intermediate optimal mean run time $\tau_f^{opt}$ (Fig. 4e). Analytically, this optimum scales with the environmental trapping time: $\tau_f^{opt} = \sqrt{\frac{\tau_{tmb}}{(\tau_t+\tau_{trp})}}\,\tau_t$ (Fig. 4f & methods).

This scaling explains why the evolved $\tau_f$ decreases with agar concentrations (Fig. 2b). A phase diagram of $\chi\,(\tau_t,\tau_f)$ highlights $\tau_f^{opt}$ (Fig. 4f), underscoring how environmental constraints shape evolutionary tuning of motility. The observed non-monotonic navigation efficiency thus emerges from a fundamental competition between enhanced diffusion and diminished bias with increasing $\tau_f$. This trade-off defines an adaptive optimum, enabling bacteria to balance exploration and directional sensing in disordered landscapes.

While this simplified model successfully captures the core trade-off for optimal chemotactic navigation, we also constructed a more rigorous framework incorporating detailed chemotaxis signaling dynamics[11,17,20,41-47]. This model integrates the adaptation dynamics of chemoreceptor free energy with the kinetics of CheY-P concentration and motor rotation, using experimentally measured constants (see Methods). From this, we derive the effective run-time deviation ($\delta\tau_R$) as: $\delta\tau_R \approx \frac{GNv_0^2}{d}\frac{\tau_{R0}'}{(1+\tau_R(F_0)/\tau)}$, where $\tau$ is adaptation time of bacterial chemotaxis signaling systems, $\tau_{R0}' \equiv \partial_F \tau_R(F_0)$ reflects sensitivity to chemoattractant gradient, $F_0$ is the baseline signaling state[20,47]. Numerical solution of the equations confirmed that the chemotactic navigation speed retains its non-monotonic dependence on $\tau_f$, validating our core theoretical insight (Fig. S7). In this detailed model, the chemotaxis bias ($B$) decreases with $\tau_f$ across biologically relevant regimes, though weak gradient sensing at very low $\tau_f$ can introduce a secondary non-monotonic trend.

The non-monotonicity of chemotactic ability $\chi$ can be understood intuitively by considering two asymptotic limits. For long $\tau_t$ ($\tau_f \ll \tau_t$), runs are primarily terminated by tumbles ($P_{trp} \ll P_{tmb}$). The bias coefficient B approximates its value in a homogeneous liquid, but the short $\tau_f$ severely limits the effective run time $\tau_R$ and thus the diffusion coefficient $D$, constraining overall performance. For short $\tau_t$ ($\tau_f \gg \tau_t$), runs are overwhelmingly interrupted by traps ($P_{trp} \gg P_{tmb}$), which impart no directional bias. Consequently, the bias coefficient $B$ — now dominated by rare tumbling events—becomes negligible. Cells thus achieve a high diffusion coefficient but an insignificant drift velocity, rendering them unable to climb gradients effectively.

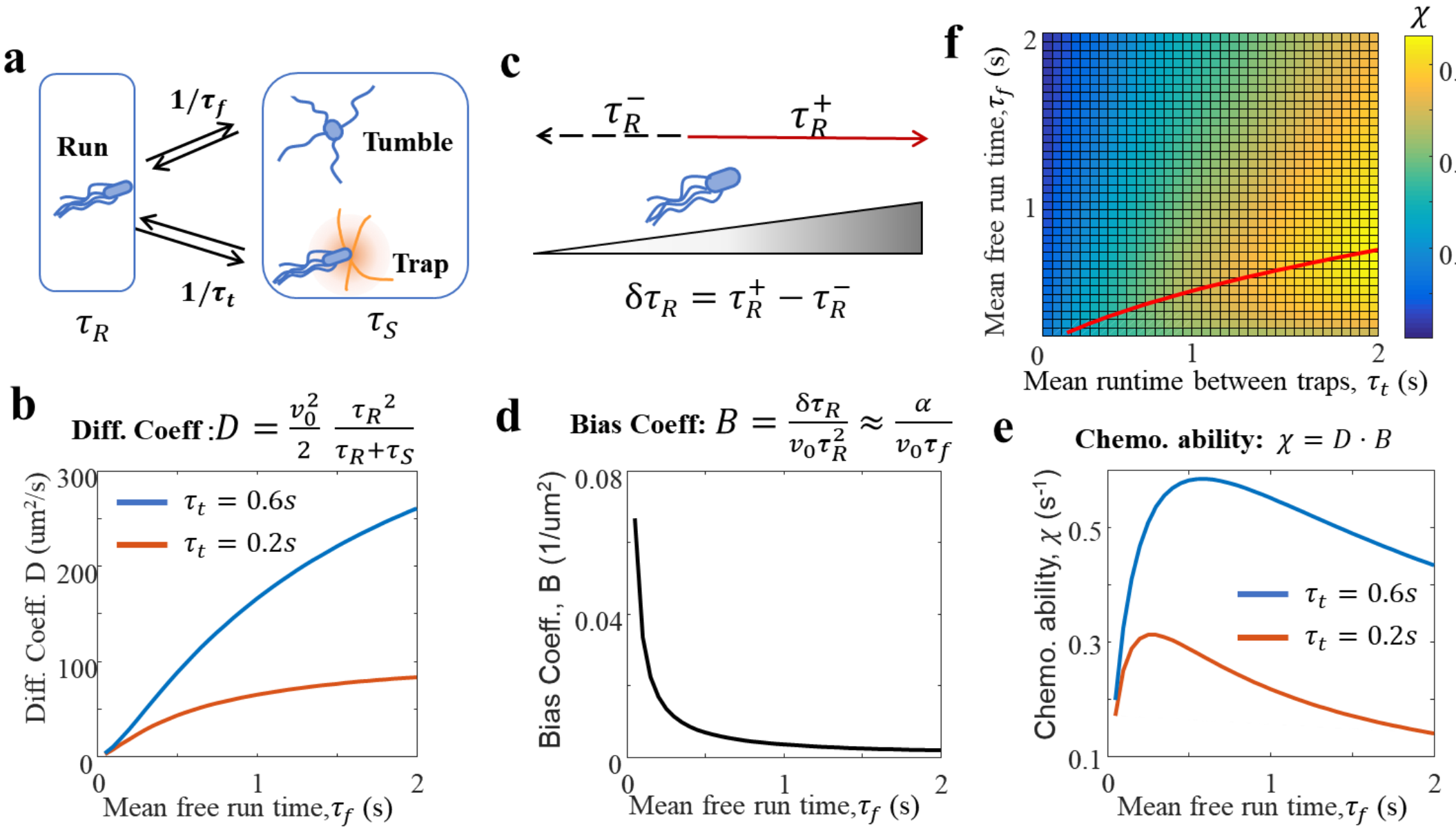

**Figure 4. A minimal model reveals a diffusion-bias trade-off.** (a) Schematic of the stochastic model combining intrinsic run-and-tumble dynamics with extrinsic trapping by the environment. (b) Predicted diffusion coefficient $D$ as a function of the intrinsic mean free run time $\tau_f$ for different mean trap intervals $\tau_t$. In the absence of a chemo-attractant gradient, $D$ increases monotonically with $\tau_f$ but is suppressed as the trap density increases (red line: low trap concentration, $\tau_t = 0.6s$; blue line: high trap concentration, $\tau_t = 0.2s$). (c) Modeling chemotaxis: cells extend runs ($\tau_f^+$) when moving up a gradient (blue) and shorten them ($\tau_f^-$) when moving down (red), creating a directional bias. (d) The bias coefficient, representing the deviation of free run durations, decreases with mean free runtime $\tau_f$. (e) the product of the diffusion coefficient and the bias coefficient yields the chemotactic ability $\chi$, which exhibits a non-monotonic dependence on the mean free runtime $\tau_f$. (f) Heatmap of chemotactic ability $\chi$

in the $(\tau_f, \tau_t)$ parameter space. The optimal mean free runtime $\tau_f^{opt}$ as a function of $\tau_t$ is shown as a red line. The external gradient $G$ was assumed to be $1\,\mu m^{-1}$ for simplicity.

## Discussion

In this study, we demonstrate that bacterial populations selected for rapid chemotaxis in agar gels adapt by tuning their mean free runtime ($\tau_f$), with distinct optimal values ( $\tau_f^{opt}$ ) emerging under different environmental constraints. Using a $\tau_f$-titratable strain, we validated that this evolutionary outcome arises from a non-monotonic relationship between the chemotactic navigation speed and $\tau_f$, where $\tau_f^{opt}$ represents a trade-off between diffusion exploration and preserved directional bias. Single-cell tracking with flagellar visualization characterized the bacterial run, trap, and tumble states and revealed that bacteria primarily escape confinement without tumbling or reorientation. By modeling chemotactic navigation as a competition between intrinsic run-tumble and extrinsic run-trap processes, we showed that the non-monotonic chemotactic ability ($\chi$), emerges from the antagonistic scaling of the diffusion coefficient (*D*) and the chemotactic bias coefficient (*B*) with $\tau_f$.

Our findings extend the understanding of bacterial navigation in porous media by examining soft and compliant gels. Whereas studies in rigid hydrogels or microfluidic pillars have highlighted "trapping-hopping" mode that may require large reorientation[24,25,48]. In the softer agar gels studied here, trap release was accompanied by only a modest change in heading, and tumbling did not appear to be the principal route out of confinement[29,37,38,49]. This difference likely stems from material differences: the flexible, transient nature of agar pores permits passive escape via mechanical yielding, whereas rigid environments may necessitate active reorientation to escape immutable traps. At the same time, tumbling is not dispensable in our system: the difference in tumbling rates up and downward gradient defines the directional bias of the run-and-tumble walk on which chemotactic navigation depends, so its role in disordered landscapes appears to be shifted from escape toward gradient sensing.

Our flagellar-state classifier performs well on bundled flagella but is less reliable for the unbundled (tumbling) state, with a mean Average Precision at an IoU threshold of 0.5 (mAP@0.5) of 0.623, and its false negatives in this class are predominantly reassigned to the bundled state (Fig. S8). Detection of tumbling events is therefore incomplete, and because these errors act in one direction, the frequency of tumbles that coincide with a trapped cell is the quantity most likely to be underestimated. Consistent with this, the 295 tumble-associated events detected among the ~9,000 trap-escape events (<3%) should be read as a lower bound on, rather than an estimate of, the true tumble fraction.

Our work underscores a fundamental distinction between active and passive particles. Passive systems obey linear fluctuation-dissipation relations ($V_D \propto D$), whereas active particles like bacteria can exhibit a non-monotonic relationship between drift velocity and diffusion coefficient. This is a direct consequence of the internal regulation of motility parameters in response to external cues. In contrast, a passive modification of bias, such as by applying an external force $G$, would yield a monotonic dependence, $V_D \propto G$.

The quantitative mismatch between the optimal free-run time $\tau_f^{opt}$ identified via titration (Fig. 2b) and the $\tau_f$ values exhibited by evolved strains (Fig. 1c) can be partially explained by the different run speeds of the strains used in the two experiments. The titration strain in Fig. 2b swims faster (~50 $\mu m/s$, Fig. S4B) than the ancestral strain used in the directed evolution experiments (~35 $\mu m/s$, Fig. S2). Within the theoretical framework, at a fixed trap density, a higher run speed reduces the mean time before a cell encounters a trap, $\tau_t$, which in turn increases the optimal mean run time as $\tau_f^{opt} = \sqrt{\frac{\tau_{tmb}}{(\tau_t+\tau_{trp})}}\,\tau_t$. Thus, the faster titration strain is expected to have a longer optimal run time (~0.9 s in Fig. 2b), whereas the slower ancestral strain should have a shorter optimum (~0.4 s in Fig. 1c). While this explanation is incomplete, the remaining quantitative deviations suggest that evolution optimizes coupled phenotypic parameters rather than isolated traits, highlighting the need for further genetic and molecular mechanistic studies.

We also noticed that the slight increase in the mean run speed during evolution (Fig. S2d) may contribute to the enhanced range expansion. Because the expansion speed scales approximately with the square root of the effective diffusion coefficient, thus scales linearly with run speed[1], so that the ~15% change we measured (Fig. S2d) is not enough to explain the >25% and >80% increases observed at 0.25% and 0.3% agar (Fig. 1b); the dominant contribution therefore comes from tuning of chemotactic sensitivity through adaptation kinetics.

By integrating evolutionary adaptation, single-cell biophysics, and theoretical modeling, we elucidate how bacteria optimize motility parameters to navigate disordered landscapes. Our findings advance the understanding of microbial ecology in porous media (e.g., soils, biofilms) and inform the design of bioengineered systems[7,8,50-52]. These results should nevertheless be read as a proof-of-principle demonstration in a well-controlled biophysical context, rather than as a universal account of bacterial evolution in nature. In particular, our evolution experiment is interpreted phenotypically: selection under controlled conditions is sufficient to test our predictions about navigation efficiency, but not to identify the underlying mutations. Genomic analysis of the evolved isolates would be required to establish that link and lies beyond the scope of this study. Extending this framework to natural habitats will therefore require testing other organisms, environmental isolates, and more complex matrices, as well as defined attractant and repellent gradients, while future work could explore how trap geometry and material elasticity modulate $\tau_f^{opt}$, and whether similar principles govern navigation in *in vivo* environments such as mucosal layers or tumor microenvironments.

**Author contribution:**

Y. Bai & X. Fu initiated this project; Y. Bai performed the data analysis and theory deduction; C. He imaged single cell behavior in agar gel, quantified the migration rate of titrated cells and performed the competition assay; W. Liu performed the directed evolution of bacteria; S. Cheng helped in single cell imaging and analysis; P. Chu helped to construct the titration strain; L. Luo helped in the theory deduction.

**Acknowledgement:**

This work is partially supported by NSFC (T2525031, 32571666, 32371492), the National Key Research and Development Program of China (2024YFA0919600, 2021YFA0910703), Strategic Priority Research Program of Chinese Academy of Sciences (XDB0480000), and General Project of Basic Research under the Shenzhen Science and Technology Plan (JCYJ20230807140810022), Natural Science Foundation of Guangdong Province, China (2025A1515011919).

**Data and code availability**

Data supporting the findings of this study are available within the main text and source files.

**Competing Interest Statement**

The authors declare no competing interests

## Materials and Methods

### Evolution protocol

Exponentially growing E. coli cells MG1655 (ancestor) were inoculated at the center of tryptone medium agar plate with 3 different agar concentrations (0.2%, 0.25%, 0.3%). After 24 hour incubation at 37 ℃, bacteria colonized the entire plate. $2\,\mu l$ cell-agar mixture was taken at the edge of the plate (25mm away from the center) and were inoculated at the center of a fresh plate of the same agar concentration. The cycle was repeated for more than 40 times. After each 10 cycles, evolved strains are saved in glycerol stocks. And the motion of the strains is tracked under microscope.

### Migration speed measurement

Semi-solid agar plates were backlit using a circular white LED array and light box, as previously described[53]. Images were captured at 30- or 60-minute intervals with a Canon EOS 600D camera and analyzed using ImageJ alongside custom MATLAB scripts. For each image, a circle was fitted to the intensity maximum to determine the colony area (A), from which the radius was calculated as $r = \sqrt{A/\pi}$. Maximum expansion speed was derived from a linear fit applied over a sliding window of at least four time points, requiring an $R^2 > 0.99$.

### Quantification of bacterial motion in liquid

To quantify phenotypic variation during evolution, evolved bacteria were cultured in tryptone broth and imaged using a 10× phase-contrast objective. For object identification, movies were background-subtracted by computing a 6-second rolling average and subtracting it from each frame within that window. Objects were then localized using the radial symmetry center method [54]. Objects were identified and tracked across frames using the U-Track software package [35].

The tracks were then separated into three states (run, tumble, or intermediate) using a previously described clustering algorithm with 3 dimensions of parameters (speed, acceleration, and angular velocity) [18]. The intermediate state represented transitions into or out of tumbles, as well as minor changes in speed and direction resulting from partial switching of flagella to clockwise rotation. More than 10,000 cells were tracked in each case, so that the statistics is creditable. The distributions and the cellular averaged values of the run time, run length, tumble times, mean speed, mean run speed are then calculated and plotted in Fig. S2 & Fig. S3.

### Chemotactic navigation and competition with CheZ titration strains

we use a synthetic strain that the free runtime is titrated by the expression level of CheZ protein. The titration of CheZ protein was released by introducing negative feedback gene circuit of *ptet-tetR* or *plac-lacI* to replace the original promotor of *cheZ* genes on chromosome as in reference[3]. In these engineered strains, CheZ expression is controlled by externally added inducers (e.g., aTc for the *ptet-tetR* system and isopropyl β-D-1-thiogalactopyranoside (IPTG) for the *plac-lacI* system), allowing fine-tuned titration of CheZ levels.

To assess the chemotactic performance of this run-time–tunable strain, semi-solid agar plates were prepared with varying concentrations of the appropriate inducer (aTc or IPTG) uniformly mixed into the agar. Identical initial cell densities of the strain were inoculated at the center of each plate to ensure consistent and comparable conditions. Bacterial chemotactic navigation was monitored over time using time-lapse camera imaging, and the chemotactic navigation was quantified by tracking the outward movement of the colony edge.

**Observation of bacterial motion in soft agar**

To understand how bacteria interact with soft agar, bacteria motions were tracked in agar gels. We first labeled the flagella of bacteria, with the method established in reference[15]. This labelled strain was then cultured in M9 minimal medium with glycerol to exponential growing phase, and then was mixed with the pre-warmed soft agar medium with corresponding concentration so that the final $OD_{600}$ was 0.04-0.06. 5 ul of this mixture was then dripped on a slide and was sealed by a cover glass. This sample was freezed in 4℃ for 5mins so that the agar solution was congealed. And then was rewarmed in 37℃ for 3mins to re-activate the bacteria before they are tracked under microscope with 60X fluorescent object. So that the flagella conformation was filmed with its position.

**Classification of bacterial motion stats**

Flagellar conformations were classified into bundled or split states using a custom-trained YOLOv5 algorithm. To develop this model, raw 16-bit images were first converted to 8-bit format in ImageJ with saturated pixels set at 0.3% to enhance contrast, after which 237 frames were extracted via Python and randomly partitioned into training (n = 208) and validation (n = 29) sets. Bacterial targets, categorized as either motile or rotating based on movement patterns, were precisely annotated along their edges using LabelImg, with all labels manually verified to ensure data reliability. Training was conducted on an NVIDIA RTX 3090 GPU with input images resized to 2016 × 2016 pixels and a batch size of 6, during which learning rates and loss function weights were dynamically adjusted to optimize parameters. Performance was evaluated through cross-validation using prediction-label consistency on the validation set as the primary metric, with iterative retraining performed until stable detection accuracy was achieved on unseen data.

To rigorously validate the reliability of our custom YOLOv5 pipeline, we employed confusion matrix analysis and precision-recall (PR) curves to assess classification accuracy across bacterial motion states. The confusion matrix summarizes classification performance by comparing predicted labels against ground truth annotations, comprising four components: true positives, false positives, true negatives, and false negatives. Additionally, the mean Average Precision at an IoU threshold of 0.5 (mAP@0.5) was employed as the primary quantitative metric for detection accuracy (Fig. S8).

Instantaneous velocity was derived from the positional data at each acquisition time point. Based on flagellar state information, bacterial motion within the agar gel was classified into three distinct states: the run state, characterized by bundled flagella and high velocity; the tumble state, marked by split flagella and low velocity; and the trap state, defined by bundled flagella accompanied by low velocity.

After identifying the swimming states of bacterial motion, we determined the run time, tumble time, and trap time for each cell in agar gel by counting consecutive segments classified as running, tumbling, or trapped. Cells with short trajectory durations (< 5 s) or maximal displacements smaller than one flagella length (< 10 μm) were excluded from the analysis, as they likely represent out-of-focus tracks or tethered cells that do not contribute meaningfully to bacterial chemotactic navigation dynamics.

To quantify reorientation during trapping and tumbling events, we segmented trajectories into distinct motifs: run–trap–run, run–tumble–run, and run–trap–tumble–run. A comparison of the frequencies of run–trap–run versus run–trap–tumble–run events revealed that cells typically escape traps without resorting to tumbling. The swimming direction of each run was defined by the vector from its start to end position. The reorientation angle was then computed as the angular difference between the directions of the runs immediately preceding and following each trap or tumble event.

**Minimal model of bacterial chemotaxis in disordered landscape**

Using the minimal model of bacterial motion in landscape with random trap (Fig. 4a), we have the effective run time $\tau_R$:

$$\tau_R = \frac{1}{\frac{1}{\tau_f} + \frac{1}{\tau_t}} = \frac{\tau_f \tau_t}{\tau_t + \tau_f}$$

We then assume that bacteria control its intrinsic free runtime $\tau_f$ by adding or deducting a portion $\alpha$ when going up or down the chemoattractant gradient $\tau_f^{\pm} = \tau_f \pm \delta\tau_f$, with $\delta\tau_f = \alpha G \tau_f$ (Fig. 4c). This assumption approximates the more realistic model integrating bacterial chemotaxis pathway. As the biasing factor is small, we can use a Taylor expansion to get the biased effective mean run time of the run-tumble particle:

$$\delta\tau_R = \left|\tau_R^{\pm} - \tau_R\right| \approx \delta\tau_f \cdot \frac{\partial\tau_R}{\partial\tau_f} = \frac{\alpha G \tau_f \tau_t^2}{\left(\tau_f + \tau_t\right)^2} = \frac{\alpha G}{\tau_f}\tau_R^2$$

The drift velocity $V_d$ is defined as:

$$V_d = \frac{v_0}{d}\frac{\tau_R^+ - \tau_R^-}{\tau_R^+ + \tau_R^- + 2\tau_S} = \frac{v_0}{d}\frac{\delta\tau_R}{\tau_R + \tau_S} \approx \frac{v_0}{d}\frac{\alpha G}{\tau_f}\frac{\tau_R^2}{\tau_R + \tau_S}$$

where $v_0$ is the mean run speed, $d$ is the dimension factor. The mean stop time $\tau_S$ is the expected duration of a stop, calculated as the sum of the tumble and trapping times weighted by their respective probabilities $P_{tmb}\tau_{tmb} + P_{trp}\tau_{trp}$. Given that a run terminates by tumbling or trapping with probabilities $P_{tmb} = \tau_f^{-1}\left(\tau_f^{-1} + \tau_t^{-1}\right)^{-1}$and $P_{trp} = \tau_t^{-1}\left(\tau_f^{-1} + \tau_t^{-1}\right)^{-1}$, thus $\tau_S$ has the form:

$$\tau_S = P_{tmb}\tau_{tmb} + P_{trp}\tau_{trp} = \frac{1/\tau_f}{1/\tau_f + 1/\tau_t}\tau_{tmb} + \frac{1/\tau_t}{1/\tau_f + 1/\tau_t}\tau_{trp}$$

Subscribing the $\tau_R, \tau_S, \delta\tau_R$, the drift velocity then simplifies to:

$$V_d \approx \frac{v_0}{d}\frac{\alpha G \tau_f \tau_t^2}{\left(\tau_f\tau_t + \tau_t\tau_{tmb} + \tau_f\tau_{trp}\right)\left(\tau_f + \tau_t\right)}$$

Plotting this drift velocity, we observe non-monotonic dependence on the intrinsic property $\tau_f$ at given disordered landscape defined by $\tau_t, \tau_{trp}$ and at fixed tumble time $\tau_{tmb}$.

This $V_d$ expression produces an Arch-shaped non-monotonic dependency on $\tau_f$, we can then calculate the optimal $\tau_f^{opt}$ that maximize $V_d$ by letting $\frac{\partial V_d}{\partial \tau_f} = 0$, then we get:

$$\tau_f^{opt} = \sqrt{\frac{\tau_{tmb}}{\left(\tau_t + \tau_{trp}\right)}}\,\tau_t$$

**Decomposition of the drift velocity**

For a run-tumble particle with no persistence of turning angles, with effective run time $\tau_R$ and effective stop time $\tau_S$, the diffusion coefficient can be written as[55]:

$$D = \frac{v_0^2}{d}\frac{{\tau_R}^2}{\tau_R + \tau_S}$$

where $\tau_R/(\tau_R+\tau_S)$ is the fraction of run time.

Revisiting the expression of chemotactic ability $\chi = \frac{V_d}{G} = \frac{1}{G}\frac{v_0}{d}\frac{\delta\tau_R}{\tau_R+\tau_S} \approx \frac{v_0}{d}\frac{\alpha}{\tau_f}\frac{\tau_R^2}{\tau_R+\tau_S}$, we note that the diffusivity D$D$ can be recovered by rearranging the terms.

$$\chi \approx \frac{v_0^2}{d}\frac{\tau_R^2}{\tau_R+\tau_S}\frac{\alpha}{\tau_f}\frac{1}{v_0} = D\cdot\frac{\alpha}{\tau_f}\frac{1}{v_0}$$

so that we denote the second term on the right side as bias coefficient

$$B = \frac{\alpha}{v_0\tau_f} = \frac{\alpha}{l_f}$$

Where $l_f$is the mean free run length of bacterial motion (Fig. S2-4). This bias coefficient is dimensionless as it represents the biased run length of a single run up and down gradient normalized by mean free run length.

**Model integrating bacterial chemotaxis signaling pathway**

The free energy of the receptor then determines the CheY-P concentration $Yp(t)$ by $Yp(t) = \frac{\alpha}{1+e^{F(t)}}$. the switching rate of the bacteria from run state to tumble state $\lambda_{rt}$ and from tumble state to run state $\lambda_{tr}$ writes $\lambda_{rt} = \omega e^{-(\frac{g}{4}-\frac{g}{2}(\frac{Yp(F)}{Yp(F)+K}))}, \lambda_{tr} = \omega e^{+(\frac{g}{4}-\frac{g}{2}(\frac{Yp(F)}{Yp(F)+K}))}$ , where $\omega, g, K$ are experimentally measured constants that describe the motor kinetics.

The bacterial receptor's free energy was adapted to an intrinsic value $F_0$ with typical adaptation time $\tau_a$, together with externally gradient $(G)$ driven modifications referenced to the chemotactic sensor $N$ and to movement relative to the gradient $\vec{r}\cdot\vec{s}v_0$.

$$\frac{dF}{dt} = -\frac{1}{\tau_a}(F-F_0) + \vec{r}\cdot\vec{s}v_0 N\, G$$

The run time in agar gel writes：

$$\tau_R = \frac{1}{\frac{1}{\tau_I}+\frac{1}{\tau_E}}$$

with

$$\tau_f = \frac{e^{\left(\frac{g}{4}-\frac{g}{2}\left(\frac{\alpha}{\alpha+K+e^{F(t)}}\right)\right)}}{\omega}$$

At shallow gradient limit where $F$ is close to its steady-state value $F_0$, one may get:

$$\tau_R^+ - \tau_R^- = \frac{2\tau_{R0}^{'} G N v_0}{\left(\frac{1}{\tau} + \frac{1}{\tau_{R0}}\right)}$$

The drift velocity writes:

$$V_d = \frac{v_0}{d} \frac{\tau_R^+ - \tau_R^-}{2\tau_{R0} + 2\tau_{S0}} \approx \frac{N v_0^2 G}{d} \frac{\tau_{R0}^{'}}{\left(1 + \frac{\tau_{R0}}{\tau}\right)} \frac{\tau_{R0}}{\tau_{R0} + \tau_{S0}}$$

These results confirm that the optimal navigation strategy of bacteria on disordered landscape requires a match between the innate free runtime $\tau_f$ and the mean free runtime between traps $\tau_t$. Cells with smaller $\tau_f$ didn't use up all the free space that the environment allows; Cells with larger $\tau_I$ have almost the same runtime whether go up or down the gradient as they are trapped to a smaller free runtime defined by $\tau_E$. This effect was more clearly illustrated by the response curve of $\tau_{R0}(F)$ as the climbing or sliding gradient modifies the internal free energy $F$ by a linear manner.

**Supplementary figures:**

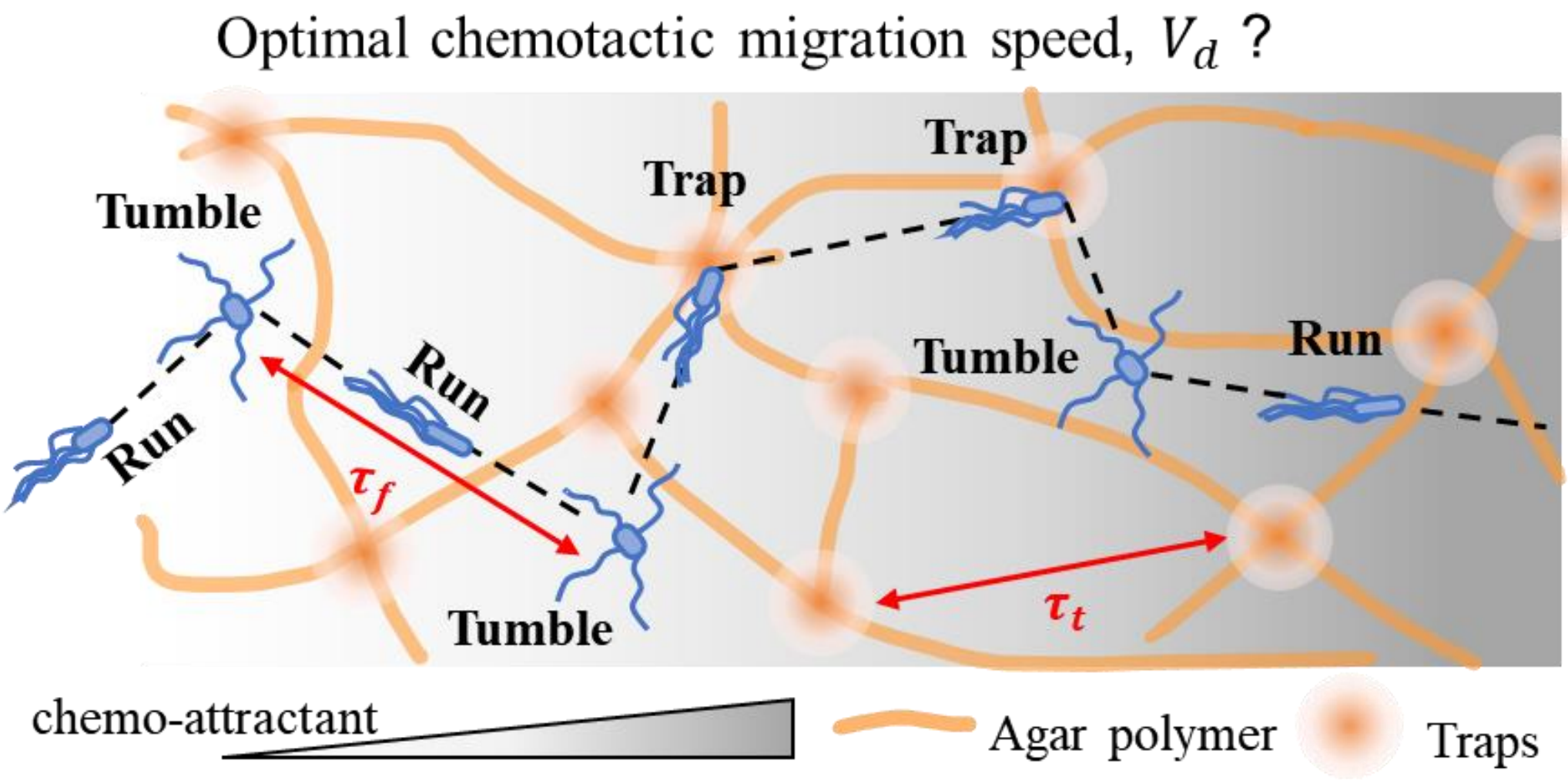

**Figure S1. Schematic illustration of bacterial chemotaxis in a porous agar gel environment.** This diagram depicts the movement of *E. coli* through a network of pores in agar gel, highlighting the three primary behavioral states: run (straight swimming with bundled flagella), tumble (reorientation via flagellar unbundling), and trap (a newly identified state where cells become temporarily immobilized due to physical confinement). The mean run duration between successive tumbles ($\tau_f$) and mean run duration between successive traps ($\tau_t$) are indicated by red arrows, representing key parameters governing motility dynamics. The orange lines represent the pore structure of the gel, while the gray background denote chemoattractant gradients. This spatially constrained environment imposes selective pressures on motility strategies of bacterial and raises questions on the optimal chemotactic navigation strategy to maximize the chemotactic navigation $V_d$ in porous agar gel.

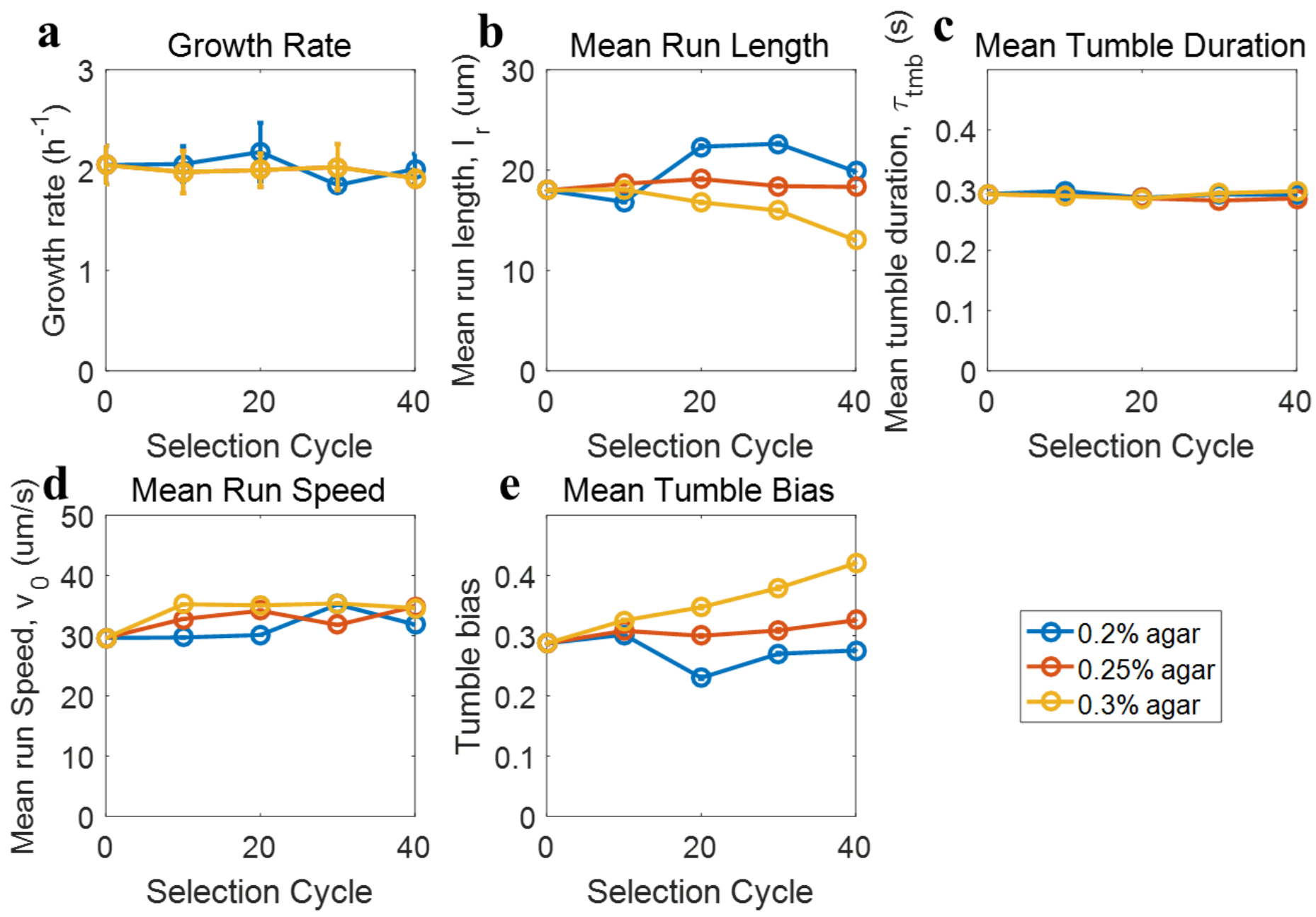

**Figure S2. Evolutionary dynamics of motility and growth parameters in liquid culture across selection cycles.** Time courses of key phenotypic traits as measured in selected E. coli populations over 40 selection cycles in three agar concentrations (0.2%, 0.25% and 0.3%). (a) Growth rates remain stable throughout the selection process, where error bars represent standard deviation of 3 independent measurements. (b) Mean run length decreases in the 0.3% agar line, while 0.25% lines remain closer to the ancestral value and 0.2% increased slightly. (c) Tumble duration maintained at a consistent level across cycles. (d) The mean run speed increases slightly (~15%) during evolution. (e) Tumble bias increases steadily in the 0.3% agar line, increases slightly in the 0.25% line, and decreases slightly in the 0.2% line. Motility related data represent averages from more than 4,800 individual cell tracks and over 100,000 run or tumble events per condition, with standard errors of the mean (SEM) smaller than the symbol size in pannel b-e.

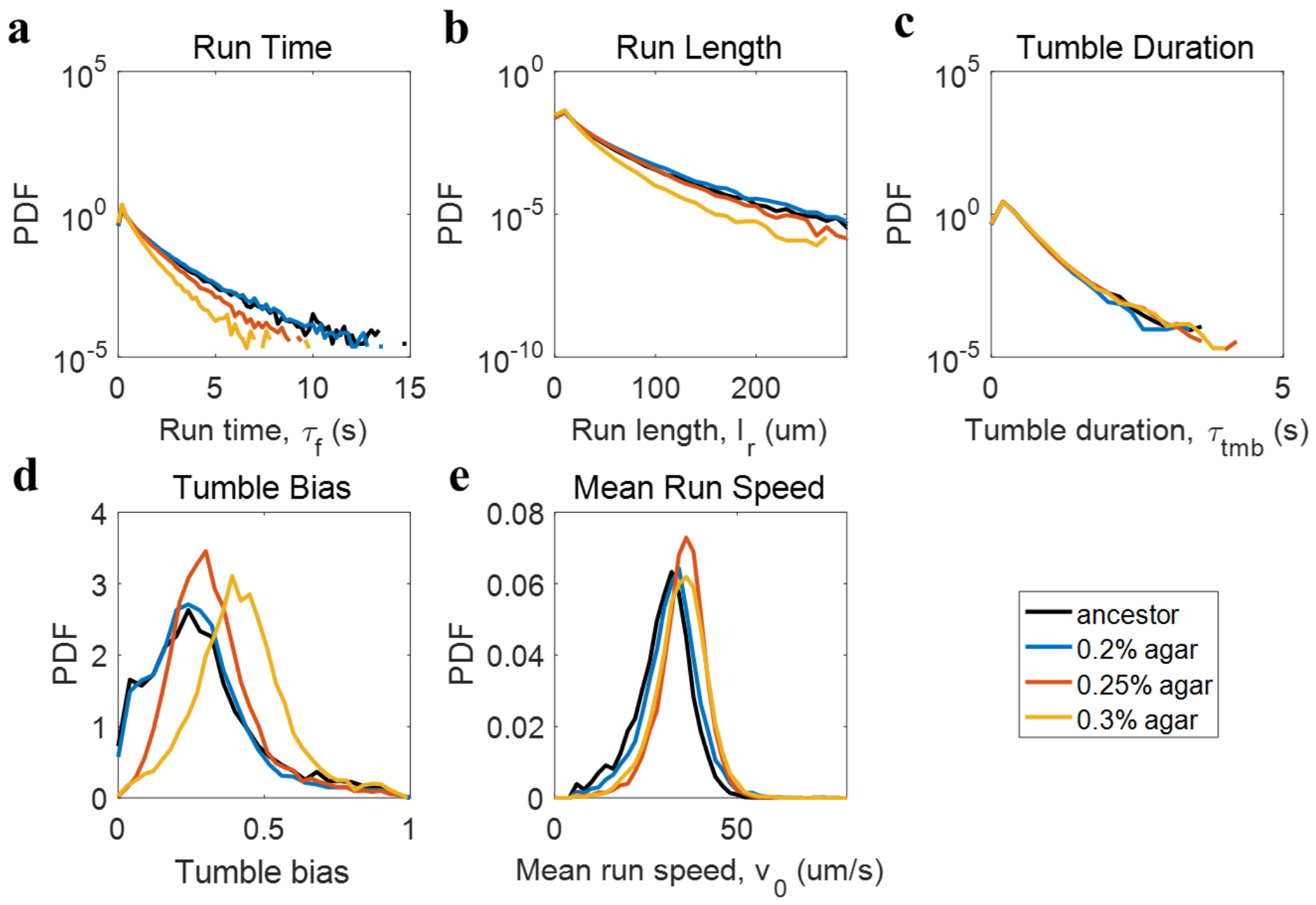

**Figure S3. Distributions of key motility parameters in evolved strains compared to the ancestral population.**
Probability density functions (PDFs) depict five fundamental motility traits measured for the ancestral strain (black lines) and three independently evolved lines selected under 0.2% (blue lines) ,0.25% (red lines) and 0.3% (yellow lines) agar concentrations, with data collected from over 100,000 run or tumble events per condition. Panels (a-c) illustrate that distributions of run times, run lengths, and tumble durations all exhibit approximately exponential decay across all strains, indicating consistent stochastic processes underlying these traits. In contrast, panels (d) and (e) show that tumble bias and mean run speed are unimodally distributed, suggesting selective pressures lead to more uniform adaptations in these parameters.

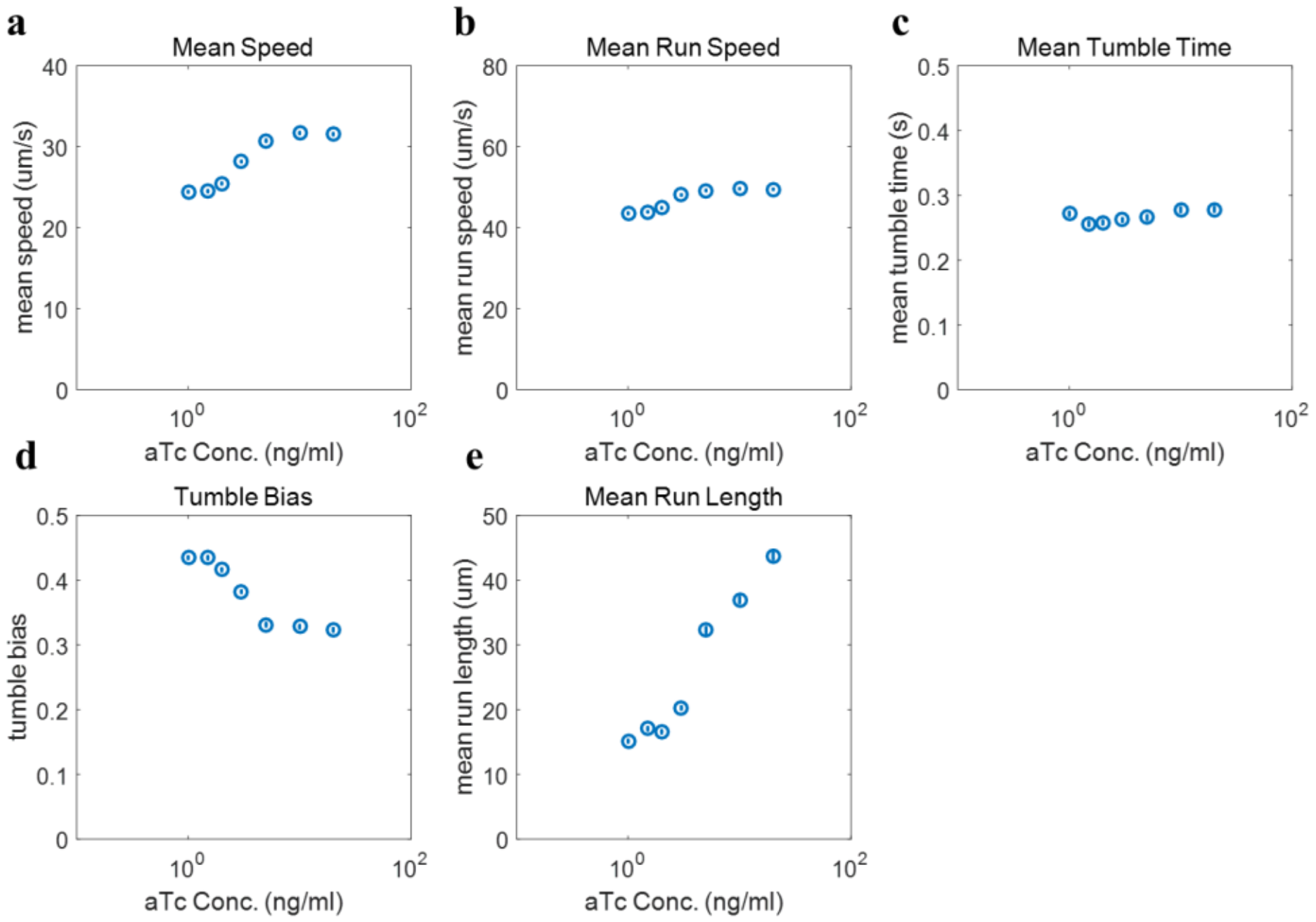

**Figure S4. aTc-dependent modulation of swimming behavior in liquid.**
(a) Mean speed of all cell tracks, including both running and tumbling events, increases with aTc concentration, indicating enhanced motor performance or reduced braking during runs. (b) Mean run speed, defined as the average speed during running events, remains approximately constant (44um/s – 50um/s) across the range of aTc concentrations tested. (c) Mean tumble time remains largely unchanged (0.25–0.28 s) across the full aTc range. (d) Tumble bias, defined as the fraction of time spent tumbling, decreases monotonically with increasing aTc, indicating increased directional persistence. (e) Mean run length increases progressively with aTc, consistent with the observed decrease in tumble bias and the increase in run time. Error bars represent the SEM; where visible, they are plotted as short lines inside the circles.

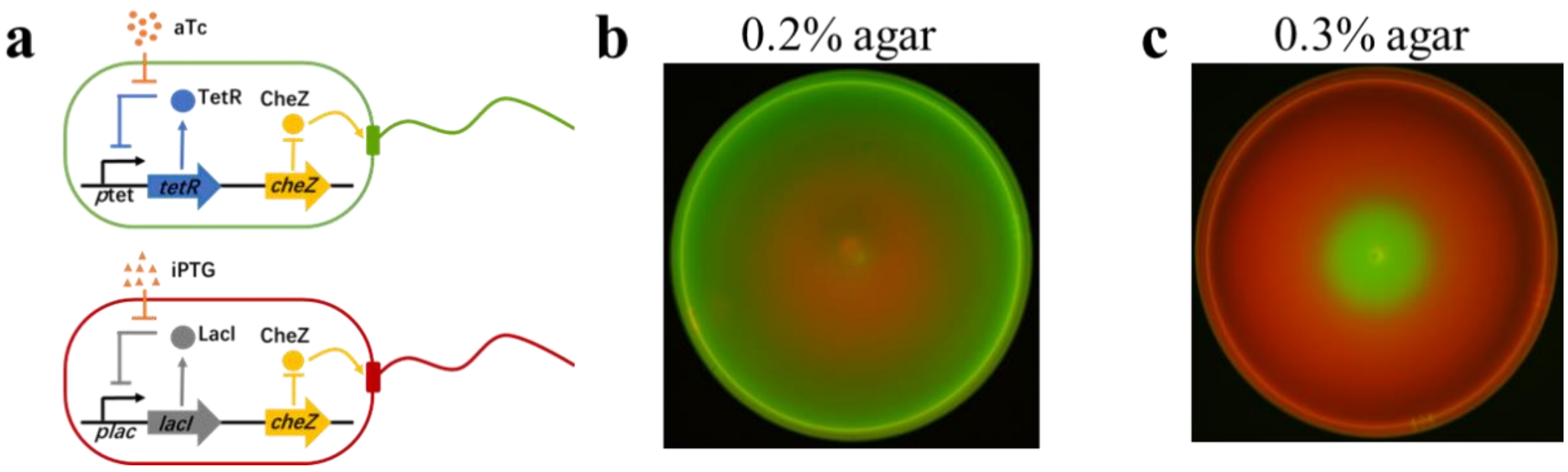

**Figure S5. Competitive fitness assay reveals environment-dependent selection of optimal run duration ($\tau_f$).** (a) Schematic representation of two genetically engineered E. coli strains, each with distinct inducible control over the mean run duration ($\tau_f$), achieved via independent expression of CheY from the tetR and lacI systems using aTc and iPTG, respectively. The strains are fluorescently labeled (green and red) for spatial tracking during competition. (b,c) Competitive range expansion assays on 0.2% agar (b) and 0.3% agar (c), where both strains were co-inoculated at equal initial density and allowed to expand overnight at 37 °C. Fluorescence imaging reveals the spatial distribution of each strain across the expanding colony. On 0.2% agar (b) the green one dominates the outer edge of the colony, indicating superior dispersal in less confined environments. In contrast, on 0.3% agar (c), this red strain is enriched toward the center expands outward, demonstrating that shorter run durations are favored under higher physical confinement.

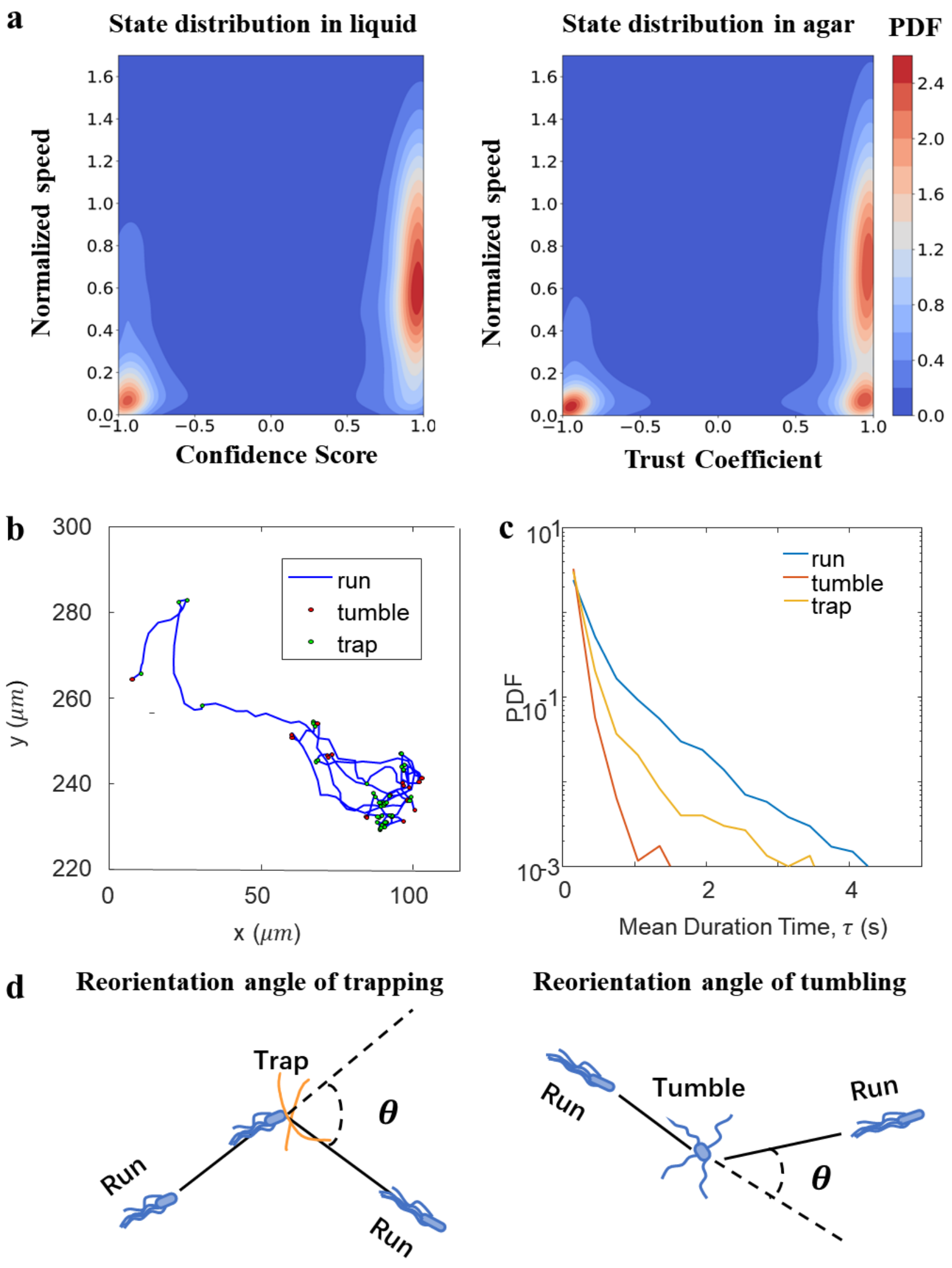

**Figure S6. Motility behavior of *E. coli* in agar gel.**

(a) Probability density distributions of cell states in agar and liquid. The confidence score represents the probability of flagellar bundling as classified by YOLOv5, ranging from −1 (fully split flagella) to 1 (fully bundled flagella). Normalized speed is calculated relative to the mean speed across all tracked trajectories. (b) Representative trajectory of a single bacterial cell moving through a 0.2% agar gel, with automatically detected behavioral states annotated: runs (blue line), tumbles (red dots), and traps (green dots). The trajectory reveals frequent reorientations and prolonged pauses indicative of physical confinement and interaction with the gel matrix. (c) Probability density functions (PDFs) of the duration for run, tumble and trap events, measured in agar gel, showing distinct temporal signatures. Runs exhibit a broad exponential decay, consistent with stochastic motility, while tumbles are brief and sharply peaked. Trap durations are longer and more variable, reflecting transient immobilization due to pore entrapment. The mean run, trapping, and tumble times are 0.30 s, 0.15 s, and 0.10 s, respectively. Distributions were plotted with more than 5,000 tumble events, more than 10,000 trapping events and 30,000 running events (d) schematic definition of reorientation angle of trapping and tumbling. $\theta$ denotes the angular change between incoming and outgoing velocity vectors during runs. Dashed lines indicate extensions of the incoming trajectories, while solid lines represent linear fits to the running trajectories.

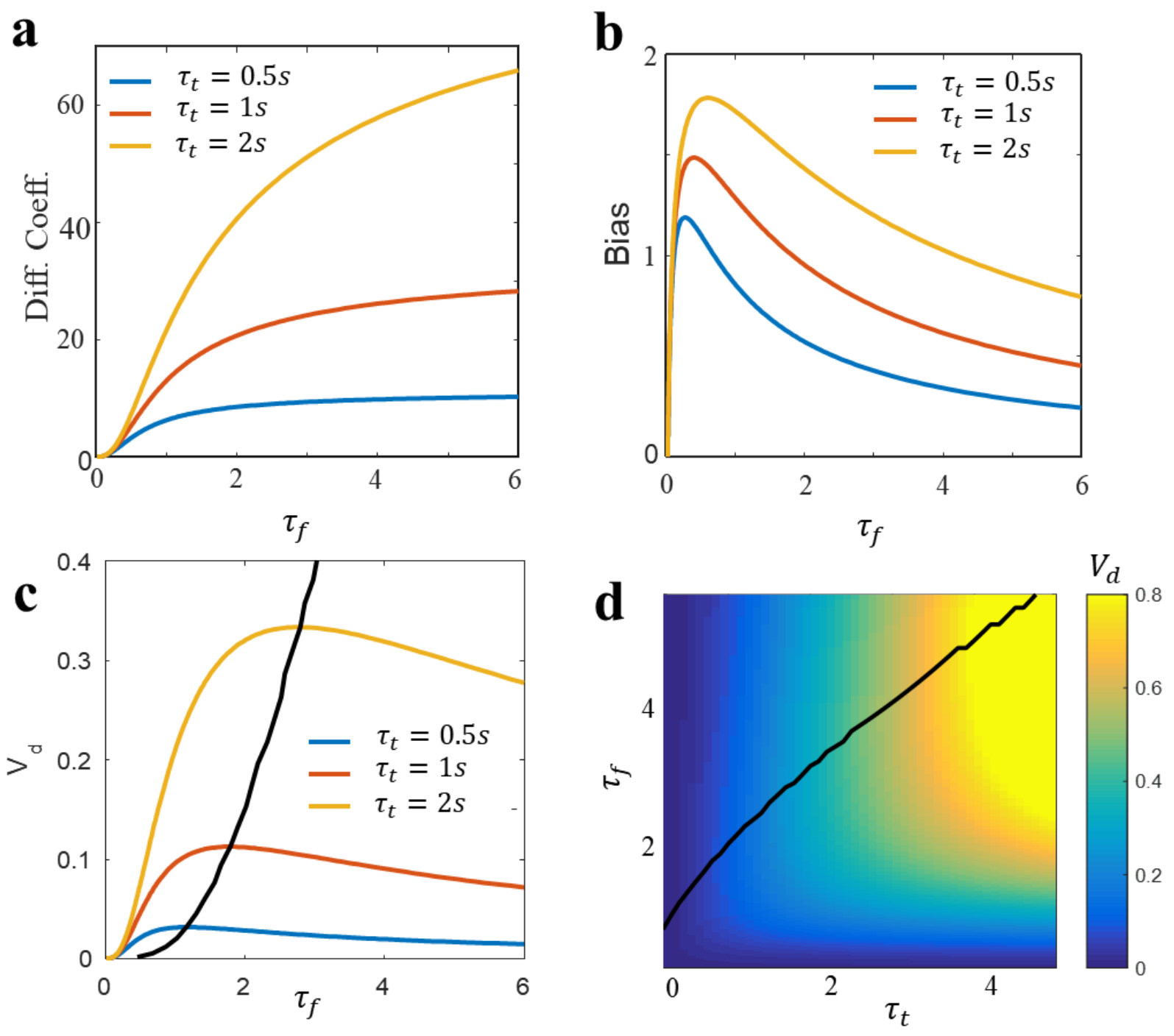

**Figure S7. Prediction of the model with chemotaxis pathway.**

Simulations incorporating the full bacterial chemotaxis network reveal how key motility metrics depend on the mean trap intervals $\tau_t$ and intrinsic run duration $\tau_f$. (a) The diffusion coefficient increases with $\tau_f$, and is further enhanced at longer $\tau_t$. (b) The chemotactic bias rises sharply with $\tau_f$ with higher initial bias at short $\tau_t$ but declines at longer $\tau_f$. (c) Effective drift velocity in a chemoattractant gradient peak at intermediate values of $\tau_f$, with optimal chemotaxis occurring when $\tau_f$ is tuned relative to $\tau_t$ (black line). (d) Contour plot of chemotactic ability ($\chi$) across a range of $\tau_f$ and $\tau_t$, revealing an increasing trend of $\tau_f^{opt}$ over $\tau_t$. This predicted dependence decreases with agar concentration, in quantitative agreement with experimentally observed behavioral tuning (Fig. 1c and Fig. 2b).

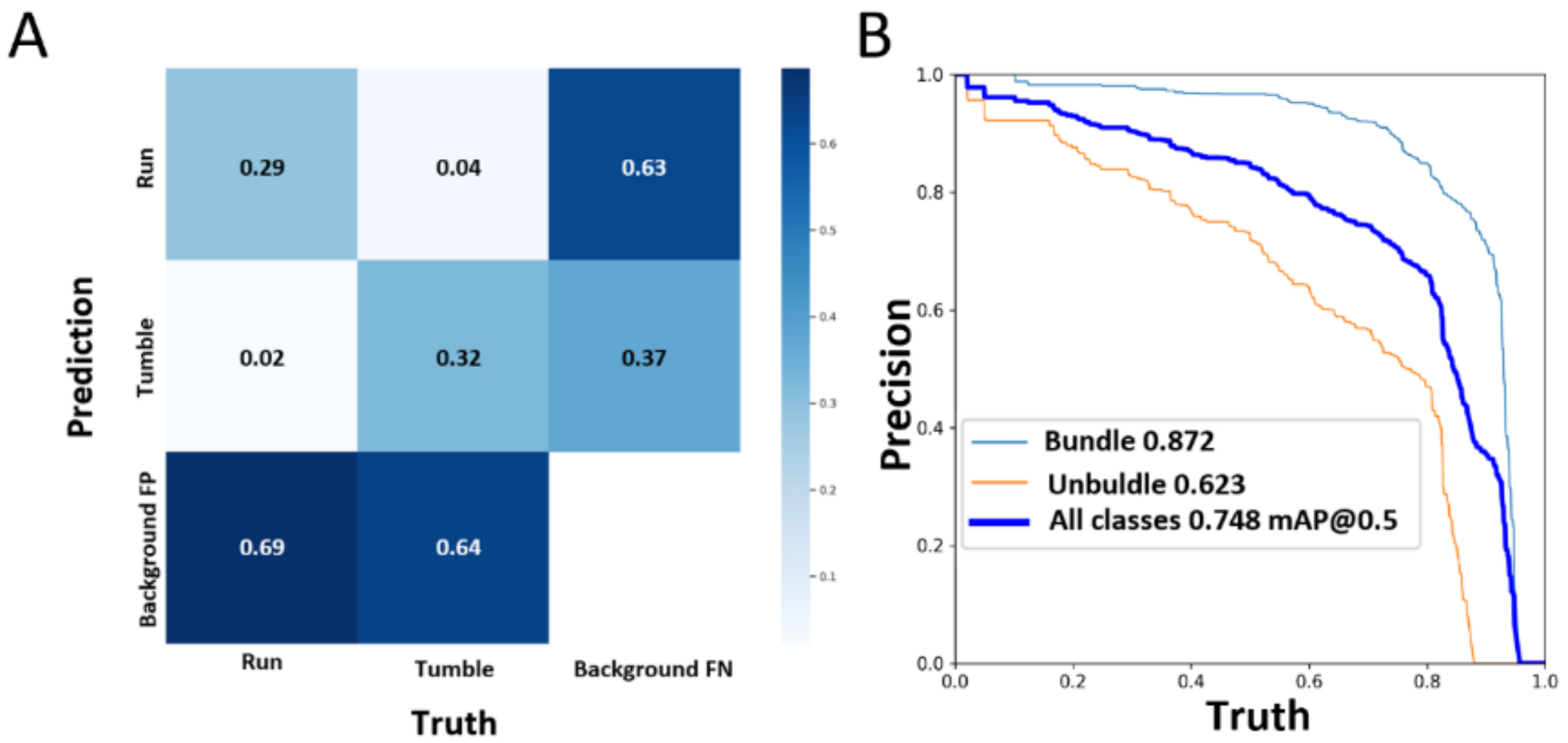

**Figure S8. Performance evaluation of the YOLOv5 model.** (a) Normalized confusion matrix illustrating the classification accuracy for bacterial flagellar "Bundle" and "Unbundle" states. Diagonal values represent the proportion of correct classifications, while off-diagonal values indicate misclassification rates. The results reveal significant confusion in identifying the "Unbundle" state, highlighting it as the primary bottleneck in current classification performance. (b) Precision-Recall (PR) curves depicting the trade-off between precision and recall across varying confidence thresholds. The mean Average Precision (mAP@0.5) for the "Bundle" class is 0.872, substantially outperforming the "Unbundle" class at 0.623, with an overall mAP@0.5 of 0.748 across all classes. Consistent with the confusion matrix analysis, the PR curves further confirm that both detection precision and recall for the unbundled flagellar state are inferior to those of the bundled state.

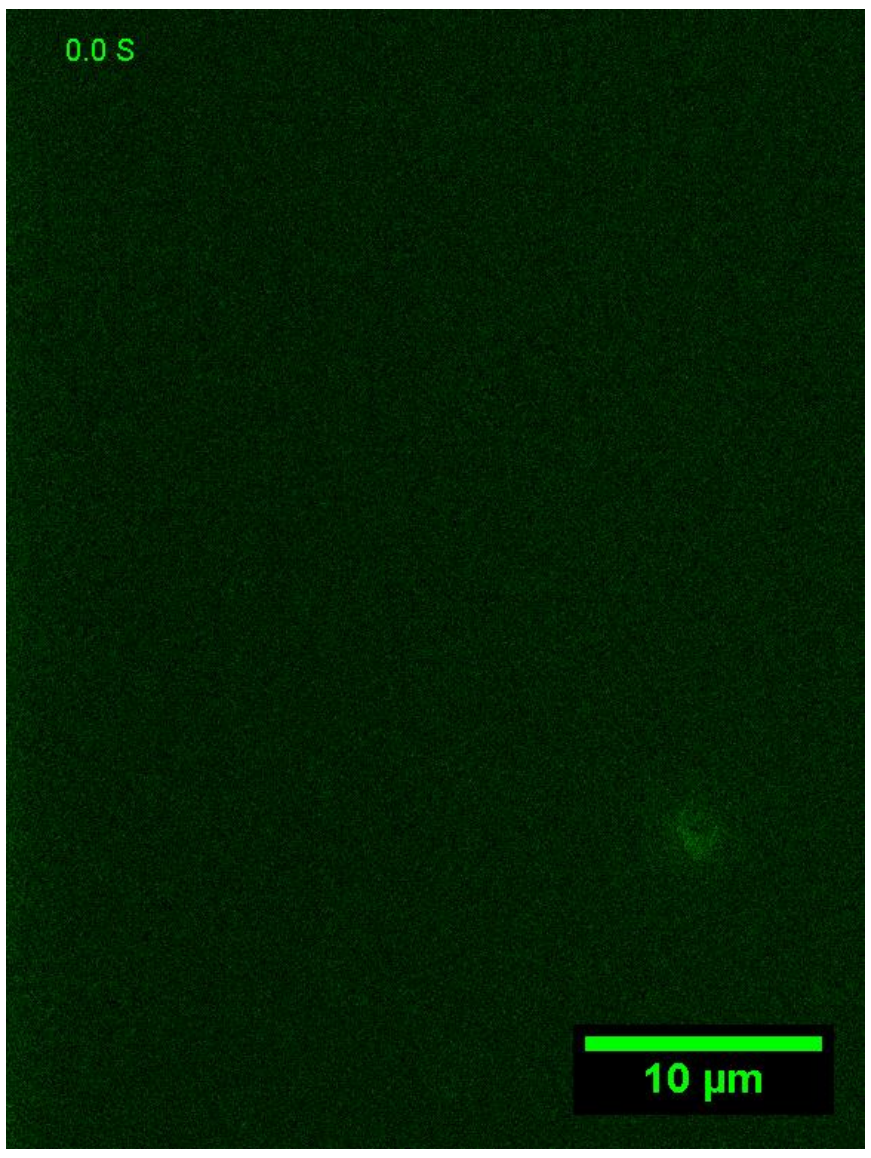

**Movie S1 Trapping state of bacteria in agar gel**

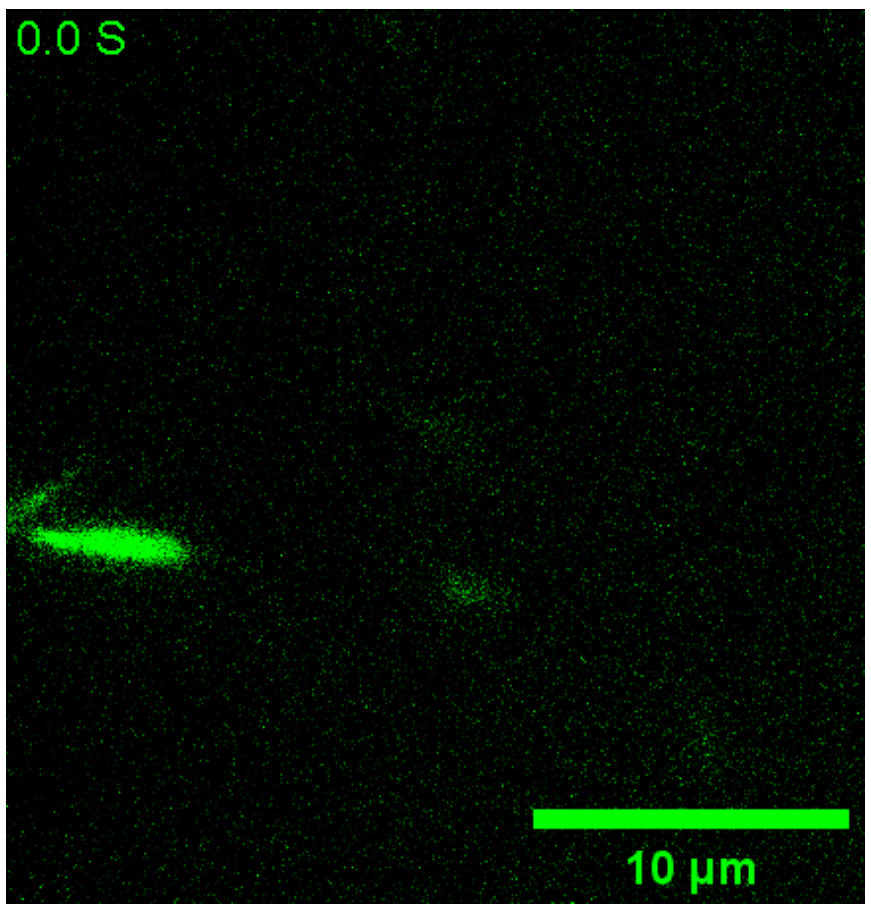

**Movie S2 Tumble state of bacteria in liquid**